\documentstyle[12pt,aps]{revtex}
\begin{document}
\title{
\begin{flushright}
\vspace{-1cm}
{\normalsize MC/TH 94/27}
\vspace{1cm}
\end{flushright}
Low-energy theorem for a composite particle in mean scalar\\ and vector fields}
\author{Michael C. Birse}
\address{Theoretical Physics Group\\
Department of Physics and Astronomy\\
The University of Manchester\\
Manchester, M13 9PL, UK\\}
\maketitle
\begin{abstract}
For a relativistic particle moving in the presence of mean scalar and vector
fields, the energy at second order in the scalar field is shown to contain two
contributions in general. One is a momentum-dependent repulsive interaction
satisfying a low-energy theorem pointed out by Wallace, Gross and Tjon. The
other does not vanish at zero-momentum and involves a ``polarisability" of the
particle by the scalar field. The first of these contributions is independent
of the details of the structure of the particle and the couplings of its
constituents to the external fields. The appearance of such a piece in the
central nucleon-nucleus potential thus would support the existence of strong
scalar fields in nuclei, without requiring the use of a Dirac equation for the
nucleon.
\end{abstract}
\vskip 20pt

Treatments of nuclei based on the Dirac equation have been very successful,
particularly in describing intermediate-energy proton-nucleus scattering
\cite{dirph,ray}. These typically involve large scalar and vector potentials of
opposite sign. Although these cancel to leave a small net central potential,
the significant reduction of the nucleon mass by the scalar potential gives
rise to a number of interesting effects. Among these are a strong momentum
dependence of the central potential and an enhanced spin-orbit coupling in
nucleon-nucleus scattering \cite{dirph,ray}, and an enhanced axial charge
\cite{axch}. For pointlike Dirac nucleons all of these arise from ``Z-graphs,"
which can be interpreted as excitations of virtual nucleon-antinucleon pairs.
However Brodsky has argued that pair creation should be suppressed by form
factors for composite systems \cite{brod} and the validity of such an
explanation for these effects has therefore been questioned.

Recently Wallace, Gross and Tjon \cite{wgt} have pointed out that, when the
Dirac equation is reduced to two-component form, the Z-graphs produce an
interaction of second-order in the scalar field that satisfies a low-energy
theorem. This interaction  provides an important momentum-dependent repulsive
piece in the central nucleon-nucleus potential. This low-energy theorem was
obtained from the classical energy of a relativistic particle and its validity
was demonstrated in a simple model for a composite fermion consisting of a
bound fermion and boson. This model was based on a zero-range force between the
constituents, and external vector and scalar fields which coupled to the
conserved vector current and scale anomaly respectively.

The appearance of the second-order interaction in an expansion of the classical
energy of a relativistic particle suggests that this result should be valid in
general. However the arguments of Ref.~\cite{wgt} relied heavily on the chosen
forms of scalar and vector couplings and left open the possibility that this
result might not apply to QCD.  The purpose of the present paper is to show
that the term identified by Wallace {\it et al.} is indeed universal and does
not depend on the details of the nucleon structure, or on how the external
fields are coupled to the constituents. I also point out that for a composite
particle in general there can be other second-order interactions, with
different momentum dependence, which do depend on its structure through
various ``polarisabilities."

Consider a composite particle moving in the presence of a uniform medium which
generates mean scalar and vector fields, $\sigma$ and $\omega^\mu$. I will
refer to the particle as a nucleon, although the results will be more general.
The dispersion relation connecting the energy and the three-momentum can be
written in the invariant form
$$G(p^2,\sigma,\omega^2,\omega\cdot p)=0. \eqno(1)$$
This form relies only on the covariance of the dynamics and it holds for both
composite and ``elementary" particles, irrespective of their spin.  It is
convenient to re-express this as an equation for $p^2$ in terms of the other
invariants,
$$p^2=F(\sigma,\omega^2,\omega\cdot p).\eqno(2)$$
For simplicity, let me assume that the $\omega$ field is sufficiently weak that
only terms to second order are needed:
$$p^2\simeq \widetilde M^2(\sigma)+[\alpha_{v}(\sigma)-G_v(\sigma)^2]\omega^2
+2G_v(\sigma)\omega\cdot p+\alpha_{p}(\sigma)(\omega\cdot p)^2. \eqno(3)$$
Here $\widetilde M(\sigma)$ is the mass of the nucleon in the absence of the
vector
field, and $G_v(\sigma)$ is its coupling to the vector field.  In general these
will depend nonlinearly on the scalar field $\sigma$. The quantities
$\alpha_{v,p}$ are ``monopole polarisabilities" of the nucleon by the uniform
vector field. A similar scalar polarisability $\alpha_s$ can be defined by
expanding $\widetilde M$ in powers of the $\sigma$ field:
$$\widetilde M(\sigma)=M_0+g_s\sigma+{\textstyle{1\over 2}}\alpha_s\sigma^2
+\cdots, \eqno(4)$$
and a mixed scalar-vector one $\alpha_{sv}$ by
$$G_v(\sigma)=g_v+\alpha_{sv}\sigma+\cdots. \eqno(5)$$
Here $M_0$ denotes the mass of a nucleon in vacuum and $g_{s,v}$ are its
couplings to the scalar and vector fields.

In the rest frame of the medium, where $\omega^\mu$ has only a time component,
Eq.~(3) can be solved to give the energy of a nucleon of momentum {\bf p}:
$$E(\hbox{\bf p})\simeq G_v\omega+\sqrt{(1+\alpha_{p}\omega^2)\hbox{\bf p}^2
+\widetilde M^2+(\alpha_{v}+\alpha_{p} \widetilde M^2)\omega^2}, \eqno(6)$$
where terms beyond $\omega^2$ have again been dropped. The rest energy of the
nucleon is
$$E^*\equiv E(\hbox{\bf p}=0)\simeq \widetilde M+G_v\omega+{\alpha_{v}
+\alpha_{p}\widetilde M^2\over 2\widetilde M} \omega^2. \eqno(7)$$
The energy for small {\bf p} can be used to define an inertial mass $M^*$ for
the nucleon by
$$E(\hbox{\bf p})\equiv E^*+{\hbox{\bf p}^2\over 2M^*}+\cdots. \eqno(8)$$
This inertial mass has the form
$$M^*\simeq \widetilde M+{\alpha_{v}-\alpha_{p} \widetilde M^2\over
2\widetilde M}\omega^2. \eqno(9)$$
Note that, because of the factor $1+\alpha_{p}\omega^2$ multiplying {\bf
p}$^2$ in Eq.~(4), the inertial mass is not, in general, equal to the Lorentz
scalar piece of $E^*$,
$$M^\prime\equiv E^*-G_v\omega\simeq \widetilde M+{\alpha_{v}+\alpha_{p}
\widetilde M^2\over 2\widetilde M}\omega^2. \eqno(10)$$

Expanding the energy to second order in the scalar and vector fields
(but keeping terms to all orders in the momentum) gives
$$E(\hbox{\bf p})\simeq \epsilon(\hbox{\bf p})+(g_v+\alpha_{sv}\sigma)\omega
+{M_0\over \epsilon(\hbox{\bf p})}\left[g_s\sigma+{\textstyle{1\over 2}}
\alpha_s\sigma^2
+{[\alpha_{v}+\alpha_{p}\epsilon(\hbox{\bf p})^2]\omega^2\over 2M_0}\right]
+{\hbox{\bf p}^2\over 2\epsilon(\hbox{\bf p})^3}g_s^2\sigma^2, \eqno(11)$$
where $\epsilon(\hbox{\bf p})$ is the energy of a free nucleon of momentum
{\bf p}. The second term in this expression is linear in $\omega$ and can be
interpreted as the vector potential experienced by the nucleon in the presence
of the fields. The third term can be expressed in terms of a scalar potential
for the nucleon,
$$S\equiv M^\prime-M\simeq g_s\sigma+{\textstyle{1\over 2}}\alpha_s\sigma^2
+(\alpha_{v}+\alpha_{p}M^2)\omega^2, \eqno(12)$$
plus a momentum-dependent piece involving the polarisability $\alpha_p$.

Finally there is the term of second order in $\sigma$ pointed out by Wallace
{\it et al.} \cite{wgt}:
$${\cal V}={\hbox{\bf p}^2\over 2\epsilon(\hbox{\bf p})^3}g_s^2\sigma^2.
\eqno(13)$$
This is a repulsive interaction which increases rapidly with the momentum of
the nucleon. Note that it involves only the scalar coupling of the nucleon as
a whole. It is thus unlike the other second-order terms which depend on the
details of the nucleon structure through the various polarisabilities. Such a
term appears in any relativistic treatment and can be thought of as arising
from the modification of the of the nucleon's mass by the scalar field. This is
clearer if we replace $g_s\sigma$ in (13) by the scalar potential $S$, which we
can do since we are considering only terms in the energy to second order in the
fields.\footnote{In fact such a replacement holds to all orders in $\sigma$
and to second order in $\omega$ for the special case of vanishing
polarisability $\alpha_{p}$.} Although the second-order term appears here as
a momentum-dependent repulsion, it is equivalent to the more familiar energy
dependence of the central potential that appears when the Dirac equation is
reduced to a two-component Schr\"odinger equation \cite{jmr}.

In general the second-order dependence of the energy (11) on the scalar field
also contains a piece which does not vanish at zero momentum. This arises from
the scalar polarisability of the nucleon $\alpha_s$ and corresponds to a
second-order dependence of the scalar potential (12) on the scalar field
$\sigma$. In the model of Ref.~\cite{wgt} the scalar field is coupled to the
scale anomaly, ensuring that the nucleon mass remains linear in $\sigma$ and so
no such term appears. (The vector polarisabilities $\alpha_{v,p}$ and
$\alpha_{sv}$ are identically zero in that model, because the vector field is
coupled to the conserved fermion current.) Similar models with more general
scalar couplings do give rise to a term of this kind in the energy \cite{bw}.

In models of the type studied by Wallace {\it et al.} the second-order
interaction ${\cal V}$ is produced by composite-fermion Z-graphs. This is a
consequence of the zero-range force between the constituents and so may not
occur in realistic treatments of nucleon structure. More generally though, such
graphs are not necessary. As has been noted in other contexts, quark
excitations and quark Z-graphs can conspire to yield the same result as nucleon
Z-graphs \cite{zgraph}. This is just what happens in the more familiar cases of
the Thomson limit of Compton scattering and low-energy theorems for $\pi$N
interactions \cite{let}, and is unsurprising given that these are all basically
classical results.

The fact that the interaction (13) does not depend on nucleon Z-graphs is
illustrated by nontopological soliton models for a nucleon embedded in mean
scalar and vector fields \cite{mmsol}, where such graphs do not appear. In
these semiclassical models, ``pushing" can be used to determine the inertial
mass of the soliton \cite{push}. For comparison with the results of
Ref.~\cite{solaxch}, the effective coupling strength of the $\omega$ to a
nucleon with zero momentum is defined by
$$g_v^*={\partial E^*\over \partial\omega}=G_v+{\alpha_{v}+\alpha_{p}
\widetilde M^2\over \widetilde M}\omega.\eqno(14)$$
The inertial mass (9) can then be expressed in the form
$$M^*=E^*-g_v^*\omega+{\alpha_{v}\over \widetilde M}\omega^2,\eqno(15)$$
to order $\omega^2$ as usual. Comparing this with Eq.~(8) of
Ref.~\cite{solaxch}, one can see that both have the same form and that the
polarisability $\alpha_{v}$ corresponds to the response of the nucleon's
structure to pushing in the presence of the vector field. The energy of the
nucleon in these models thus has the form discussed above and contains a
piece of the form (13).

Although such a momentum-dependent repulsion could provide good evidence for
strong scalar fields in nuclei, it has proved difficult to identify
unambiguously such a term in the nucleon-nucleus optical potential
\cite{ray,kfmr}. As well as the mean-field effects included in (1), the
self-energy of a nucleon in matter should include Pauli-exchange (Fock) terms.
These are nonlocal and can lead to a similar momentum dependence to that
generated by the relativistic effect studied here. As discussed by Kleinmann
{\it et al.} \cite{kfmr}, it is possible that, at least at low energies, the
large Fock terms required for non-relativistic parametrisations of the NN
interaction are mocking up the momentum dependence of a relativistic
description.

Various other properties of a nucleon in the nuclear medium are sensitive to
the reduction in mass caused by the scalar potential. Of particular interest is
the axial charge whose observed enhancement \cite{warb} is too large to be
explained by pion exchange effects and so provides strong evidence for scalar
fields in nuclei \cite{axch}. Nuclear magnetic moments are also influenced by
these fields but they do not provide an unambiguous signal. Spin dependent
couplings to external axial or magnetic fields can be included in (1) and used
to extract effective coupling constants. However all of these involve
polarisabilities of the nucleon and so do not satisfy low-energy theorems.
This is illustrated by soliton and bag model calculations of axial couplings
and magnetic moments in medium \cite{mmsol,solaxch,saito}, where the results
depend on the details of the nucleon structure. The only quantities which do
satisfy such theorems are the orbital magnetic $g$-factors. These arise from
the part of the current that classically is proportional to the nucleon
velocity, and hence is inversely related to its mass. The enhancement of these
due to a mean scalar field is independent of the details of the nucleon
structure. Unfortunately  there are too many other exchange-current and
configuration-mixing  contributions to the orbital $g$-factors to allow to
allow this enhancement to be identified in measured magnetic moments
\cite{trb}.

In summary: for a relativistic particle in the presence of scalar and vector
fields the momentum-dependent repulsive interaction of second order in the
scalar potential is universal. The appearance of such behaviour in the central
nucleon-nucleus potential could thus provide evidence for strong scalar fields
in nuclei, independently of the use of a Dirac equation for the nucleon.

\section*{Acknowledgments}
I am grateful to S. J. Wallace for extensive correspondence on these ideas.
This work was supported by the EPSRC.


\begin{references}
\bibitem{dirph}S. J. Wallace, 1987 Annu.\ Rev.\ Nucl.\ Part.\ Sci.\ {\bf 37},
267 (1987).
\bibitem{ray}L. Ray, G. W. Hoffmann and W. R. Coker, Phys.\ Reports
{\bf 212}, 223 (1993).
\bibitem{axch} J. Delorme and I. S. Towner, Nucl. Phys. {\bf A475}, 720 (1987);
M. Kirchbach, D. O. Riska and K. Tsushima, Nucl. Phys. {\bf A542}, 616 (1992);
I. S. Towner, Nucl. Phys. {\bf A542}, 631 (1992).
\bibitem{brod}S. J. Brodsky, Comments Nucl.\ Part.\ Phys.\ {\bf 12}, 213
(1984).
\bibitem{wgt}S. J. Wallace, F. Gross and J. A. Tjon, University of Maryland
preprint, nucl-th/9407040 (1994), Phys.\ Rev.\ Lett.\ (to be published).
\bibitem{jmr}M. Jaminon, C. Mahaux and P. Rochus, Phys.\ Rev.\ {\bf C22},
2027 (1980).
\bibitem{bw}M. C. Birse and S. J. Wallace, unpublished.
\bibitem{zgraph}B. K. Jennings, Phys.\ Lett.\ {\bf B196}, 307 (1987);
J.~Achtzehnter and L. Wilets, Phys.\ Rev.\ {\bf C38}, 5 (1988);
T. Jaroszewicz and S. J. Brodsky Phys.\ Rev.\ {\bf C43}, 1946 (1991);
T. D. Cohen, Phys.\ Rev.\ {\bf C45}, 833 (1992).
\bibitem{let}J. A. McGovern and M. C. Birse, Phys.\ Rev.\ {\bf D49}, 399
(1994).
\bibitem{mmsol}M. K. Banerjee, Phys.\ Rev.\ {\bf C45}, 359 (1992);
E. Naar and M. C. Birse, J.\ Phys. G {\bf 19}, 555 (1993).
\bibitem{push} P. Ring and P. Schuck, {\it The nuclear many-body problem},
(Springer-Verlag, New York, 1980);
M. C. Birse, Prog. Part. Nucl. Phys. {\bf 25}, 1 (1990).
\bibitem{solaxch}M. C. Birse, Phys.\ Lett.\ {\bf B316}, 472 (1993).
\bibitem{kfmr}M. Kleinmann, R. Fritz, H. M\"uther and A. Ramos, University
of T\"ubingen preprint, nucl-th/9402025 (1994).
\bibitem{warb}E. K. Warburton, Phys.\ Rev.\ Lett.\ {\bf 66}, 1823 (1991);
Phys. Rev. {\bf C44}, 233 (1991);
E. K. Warburton and I. S. Towner, Phys.\ Lett.\ {\bf B294}, 1 (1992).
\bibitem{saito}K. Saito and A. W. Thomas, University of Adelaide preprint,
ADP-94-20/T160, nucl-th/9410031 (1994).
\bibitem{trb}K. Tsushima, D. O. Riska and P. G. Blunden, Nucl.\ Phys.\ {\bf
A559}, 543 (1993).
\end{references}
\end{document}